\documentclass[aps,prl,twocolumn,showpacs,superscriptaddress,groupedaddress]{revtex4}

\usepackage{amsmath,amsfonts}
\usepackage{bm}% bold math
\usepackage{graphicx}
\usepackage[pdftex,usenames,dvipsnames]{color}
\usepackage{comment}

\def\opone{\leavevmode\hbox{\small1\kern-3.8pt\normalsize1}}

\begin{document}
\title{Optical decoherence studies of Tm$^{3+}$:Y$_3$Ga$_5$O$_{12}$}
\author{C. W. Thiel}
\affiliation{Department of Physics, Montana State University, Bozeman, Montana 59717, USA}
\author{N. Sinclair}
\affiliation{Institute for Quantum Science and Technology, and Department of Physics \& Astronomy, University of Calgary, Calgary, Alberta T2N 1N4, Canada}
\author{W. Tittel}
\affiliation{Institute for Quantum Science and Technology, and Department of Physics \& Astronomy, University of Calgary, Calgary, Alberta T2N 1N4, Canada}
\author{R. L. Cone}
\affiliation{Department of Physics, Montana State University, Bozeman, Montana 59717, USA}

\begin{abstract}
Decoherence of the 795 nm $^3$H$_6$ to $^3$H$_4$ transition in 1\%Tm$^{3+}$:Y$_3$Ga$_5$O$_{12}$ (Tm:YGG) is studied at temperatures as low as 1.2 K. The temperature, magnetic field, frequency, and time-scale (spectral diffusion) dependence of the optical coherence lifetime is measured. Our results show that the coherence lifetime is impacted less by spectral diffusion than other known thulium-doped materials. Photon echo excitation and spectral hole burning methods reveal uniform decoherence properties and the possibility to produce full transparency for persistent spectral holes across the entire 56 GHz inhomogeneous bandwidth of the optical transition. Temperature-dependent decoherence is well described by elastic Raman scattering of phonons with an additional weaker component that may arise from a low density of glass-like dynamic disorder modes (two-level systems). Analysis of the observed behavior suggests that an optical coherence lifetime approaching one millisecond may be possible in this system at temperatures below 1 K for crystals grown with optimized properties. Overall, we find that Tm:YGG has superior decoherence properties compared to other Tm-doped crystals and is a promising candidate for applications that rely on long coherence lifetimes, such as optical quantum memories and photonic signal processing.
\end{abstract}
 

%Optical transient phenomena: quantum beats, photon echo, free-induction decay, dephasings and revivals, optical nutation, and self-induced transparency
%Spectroscopy of solid state dynamics
% Rare-earth ions and impurities
%Coherent nonlinear optical spectroscopy
%03.67.Hk Quantum communication
%78.47.N High resolution nonlinear optical spectroscopy

\maketitle

\section{Introduction}

The performance of many coherent signal processing applications is limited by decoherence \cite{mandel1995a}. Among the most promising materials offering long optical and spin coherence lifetimes are cryogenically-cooled rare-earth-ion-doped crystals \cite{macfarlane1987a,sun2005a,tittel2010a,thiel2011a}. The optically-addressable 4f$^N$-4f$^N$ electronic transitions in these materials are only weakly affected by the crystal environment, allowing slow decoherence and many other properties approaching that of isolated atoms. Due to these unique properties, rare-earth-ion-doped crystals have been employed in a wide variety of photonic signal processing demonstrations that include radio-frequency signal processing (see Ref. \cite{babbitt2014a} and references therein), holographic optical memories \cite{renn2002a,schlottau2004a}, high-resolution ultrasound imaging \cite{li2008a,tay2010a}, laser frequency stabilization \cite{strickland2000a,thorpe2011a,chen2011a,thiel2014a}, optical quantum memories \cite{tittel2010a,lvovsky2009a,bussieres2013a}, and many others.

Recent studies by Thiel et al. \cite{thiel2014b} have found 1\%Tm$^{3+}$:Y$_3$Ga$_5$O$_{12}$ (Tm:YGG) to have superior optical decoherence properties compared to known Tm$^{3+}$-doped and many other rare-earth-ion-doped materials. While Tm:YGG has an absorption strength that is weaker than other Tm$^{3+}$-doped materials and only somewhat larger than Eu$^{3+}$ and Tb$^{3+}$ materials \cite{thiel2011a}, the absorption strength can be enhanced for applications by utilizing impedance-matched cavities if necessary \cite{afzelius2010b,moiseev2010a}. Motivated by the potential to provide a suitable combination of properties needed for some demanding photonic and quantum applications, we performed a detailed study of decoherence of the inhomogeneously broadened lowest $^3$H$_6$ to $^3$H$_4$ transition in Tm:YGG at temperatures as low as 1.2 K.

Decoherence in rare-earth-ion-doped materials can be caused by a number of physical mechanisms, including phonon interactions, fluctuating electromagnetic fields arising from random nuclear and electronic spin-flips in the host crystal, and dynamic disorder modes in the host lattice structure \cite{macfarlane1987a,sun2005a}. To elucidate these processes in Tm:YGG, we employ two- and three-pulse photon echo measurements to measure the temperature and magnetic field dependence of the optical coherence lifetime as well as time-dependent broadening caused by spectral diffusion phenomena. Moreover, spectral hole burning and photon echoes are used to show uniform decoherence properties as a function of wavelength, verifying that all Tm$^{3+}$ ions in the lattice experience the same decoherence dynamics. Our results show that at temperatures less than 1 K, refined crystal growth, and possible optimization of crystal orientations, coherence lifetimes approaching one millisecond may be possible in this system.

\section{Tm:YGG crystal properties}

The crystal Y$_3$Ga$_5$O$_{12}$ (YGG) has cubic symmetry described by space group Ia3d with eight formula units per unit cell \cite{menzer1928a}. The Y$^{3+}$ ions occupy crystallographically equivalent lattice sites with dodecahedral point symmetry (D$_2$ point group) that have six different local site orientations related to each other through the overall cubic symmetry of the crystal \cite{dillon1961a}. Trivalent rare-earth ions substitute for Y$^{3+}$ in the lattice without charge compensation and the stoichiometric concentration of Y$^{3+}$ sites is 1.298$\times$10$^{22}$ ions$/$cm$^3$. The transition between the lowest energy components of the $^3$H$_6$  and $^3$H$_4$ multiplets occurs at a vacuum wavelength of 795.325 nm in this material, and single crystal YGG has a predicted isotropic index of refraction of $n=1.95$ at this wavelength, \cite{wemple1973a} a value that we confirm using an optical reflectometer. While offering unusually long excited-state and coherence lifetimes of 1.3 and 0.49 milliseconds respectively \cite{thiel2014b}, this transition exhibits a weaker oscillator strength in this material than found in other leading Tm$^{3+}$-doped systems \cite{sun2005a,thiel2014b}. The relatively small absorption is an apparent consequence of the transition only being weakly allowed due to small perturbations of the ideal D$_2$ point symmetry \cite{sun2005a}, likely caused by the unusually high density of anti-site defects that occur in the garnets \cite{brandle1974a,dong1991a,lupei1995a}. Spectral diffusion in Tm:YGG is expected to arise from spin flips of the gallium and yttrium nuclear spins in the host lattice. Yttrium has a single spin $1/2$ isotope $^{89}$Y with a free-nucleus gyromagnetic ratio of 2.1 MHz/T and natural abundance of 100\% \cite{lee1967a}. Gallium has two spin 3/2 isotopes $^{69}$Ga and $^{71}$Ga with free-nucleus gyromagnetic ratios of 10.2 MHz/T and 13.0 MHz/T, and natural abundance of 60\% and 40\%, respectively \cite{lee1967a}.

\section{Experimental methods}

The samples are mounted in an Oxford Optistat liquid helium cryostat. For temperatures below 2.17 K, the samples are immersed in superfluid liquid helium and the temperature is determined by measuring the vapor pressure of the liquid. For higher temperatures, the samples are cooled by a constant flow of helium exchange gas with the temperature of the gas measured using a Rh-Fe resistance sensor in the cryostat. Homogeneous magnetic fields of up to $\sim$ 500 G are applied using a water-cooled Helmholtz coil. A higher field of 6.4 kG is generated by mounting the crystal between a pair of N50-grade NdFeB block magnets immersed in the liquid helium. The magnetic field strength is determined from room temperature measurements corrected for the temperature dependence of the NdFeB magnetic field at $\sim$ 2 K. A Coherent 899-21 Ti:Sapphire ring laser is used as the light source with an estimated linewidth of $\sim$ 100 kHz and output power of typically 425 mW. The frequency is monitored using a Burleigh WA-1500 wavemeter with absolute accuracy of better than 1 GHz and a relative precision of better than 100 MHz.  The maximum laser power at the sample is typically 100 mW for coherent transient measurements and $<$ 1 mW for hole burning.

A pair of acousto-optic modulators (AOMs) in series gates the laser beam to generate pulses for photon echo and hole burning measurements. Spectral holes are burned and probed by ramping the laser frequency using a third double-passed AOM driven by a voltage-controlled-oscillator. Another AOM is placed before the detector to block excitation pulses and selectively pass emitted echo signals. In this configuration, the measured on/off dynamic range of the generated pulses is $>$80 dB and the extinction of the gating AOM is typically 40 dB. 

Optical transmission is detected using a New Focus 1801 amplified silicon photodiode. To detect photon echoes, a Hamamatsu R928 photomultiplier with extensive light baffles is used with a voltage of -1250 V. When measuring weaker echo signals, the photomultiplier termination is chosen to be 1-10 k$\Omega$ to maximize the voltage gain. Echo signals are digitized, the background light level subtracted, and the signal is integrated over time to measure the total power emitted in the echo. The voltage gain of the detection is automatically adjusted at each data point to maximize the sensitivity and effectively provide logarithmic amplification of the signal. 

We employ pulse lengths of 200 ns for all pulses in two-pulse and three-pulse echo measurements to provide sufficient echo signal strength while maximizing the pulse bandwidth to reduce the impact of laser frequency noise. Due to persistence, efforts are made to minimize hole burning and accumulated echo distortions of the observed signals whenever a magnetic field is applied to the crystal. This included continuously scanning the laser frequency so that different regions of the inhomogeneous line are sampled on each shot, with typical continuous laser scan ranges during echo measurements of 500 MHz over 625 seconds.

\begin{figure}
\begin{center}
\includegraphics[width=\columnwidth]{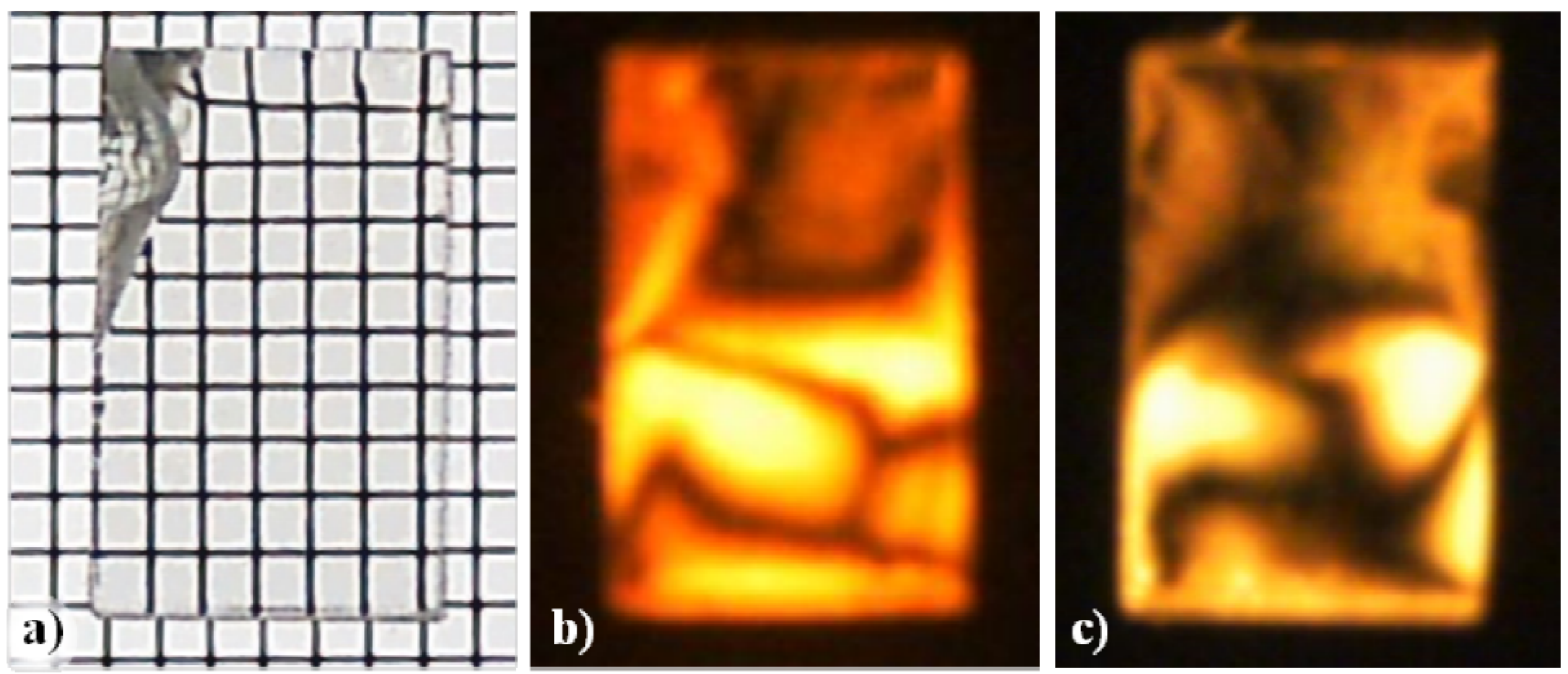}
\caption{
a) Picture of a 1 mm square grid pattern viewed through the 10 mm thick 1\%Tm:YGG crystal used for all measurements. b) Strain-induced birefringence observed throughout the crystal using a polariscope.  c) Birefringence pattern with the crossed polarizers rotated 90 degrees relative to (b).
} 
\label{fig:birefringence}
\end{center}
\end{figure}

Measurements are carried out on a 10 mm-thick single crystal of Tm:YGG from Scientific Materials Corp. (growth number 4-223). For all measurements, the laser propagates along a $<$110$>$-direction in the crystal, and the magnetic field and linear optical polarization are parallel to the orthogonal $<$111$>$ axis. While polishing the sample, we observed a noticeable tendency for the surfaces to chip, a potential indication of large internal strains in the crystal. To further investigate this, we examined the crystal using a polariscope and observed unusually large inhomogeneous strain-induced birefringence throughout the sample. See Fig. \ref{fig:birefringence} for images of the Tm:YGG crystal taken with unpolarized light (a) and illuminated in a linear polariscope for two different orientations of the crossed polarizers (b) and (c). While an isotropic crystal would normally appear dark, the presence of large strain-induced birefringence produces the patterns of light and dark fringes shown in Fig. \ref{fig:birefringence}. Similar birefringence was found for other samples obtained from the boule, suggesting the strain is a result of the growth process and not the details of the sample preparation. Otherwise, the sample was transparent and colorless with no other apparent defects observed optically and in the Laue x-ray diffraction pattern.

\section{Results}

\subsection{Frequency dependence}
 
\begin{figure}
\begin{center}
\includegraphics[width=\columnwidth]{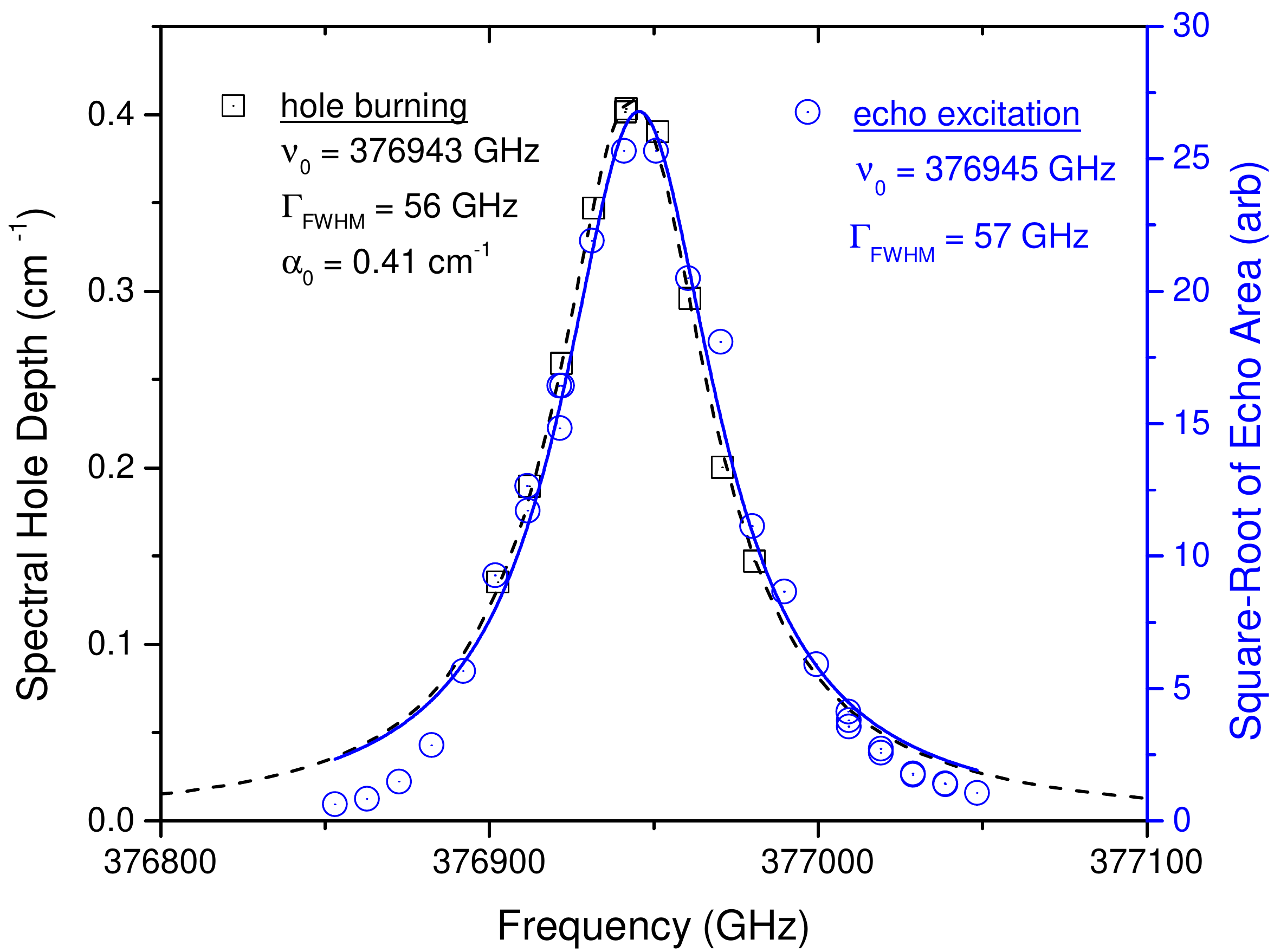}
\caption{Spectral hole burning and photon-echo-excitation spectra of the inhomogenous absorption line. The photon echo excitation spectrum (circles) closely matches the lineshape determined using spectral hole burning (squares), indicating that the entire inhomogeneous line has similar decoherence properties. Parameters extracted from fits to echo excitation spectra (solid line) and hole burning (dashed line) are shown.
} 
 \label{fig:lineshape}
\end{center}
\end{figure}

Previous broadband absorption measurements of the 795 nm transition of Tm:YGG found an estimated full-width at half-maximum linewidth of 55 GHz and peak absorption of 0.53 cm$^{-1}$ after deconvolving the spectrometer resolution from the measured spectrum \cite{sun2005a}. The linewidth is more than a factor of two larger than that of the closely related Tm$^{3+}$:YAG material, an indication of an increase in random nano-scale lattice strains potentially caused by a higher density of Y-Ga anti-site defects in YGG than Y-Al anti-site defects in YAG \cite{dong1991a,stanek2013a}. As an initial step in evaluating whether the increased inhomogeneous bandwidth may be fully employed and to determine if any potential underlying “parasitic” absorption is present due to highly perturbed Tm$^{3+}$ sites in the lattice, we perform spectral hole burning across the entire absorption line. In these measurements, optical pumping by the excitation laser was used to generate complete transparency at each optical frequency across the absorption line. The observed variation in maximum hole depth is recorded as a function of frequency, and is shown in Fig. \ref{fig:lineshape}. The resulting lineshape is fit well by a Lorentzian with FWHM of 56 GHz and peak absorption coefficient of 0.41 cm$^{-1}$, giving reasonable agreement with the broadband absorption measurements reported in \cite{sun2005a}. The center of the absorption line is found to be 376.943 THz, or 795.325 nm in vacuum \cite{thiel2014b}.
   
To further probe the underlying decoherence dynamics across the entire inhomogeneous line, we measure two-pulse photon-echo-excitation spectra \cite{sun2012a}. For these measurements, two excitation pulses with a fixed relative delay of $t_{12}$ are used to first prepare a coherent superposition of the ground and excited electronic states for the ensemble of resonant Tm$^{3+}$ ions in the crystal, and then to rephase the inhomogneous-broadening-induced dephasing and subsequently produce a coherent burst of radiation, or photon echo, after an additional delay of $t_{12}$. The integrated power of the emitted photon echo is monitored as the frequency of the laser is varied across the absorption line. The echo power $I_{echo}$ is extremely sensitive to the specific decoherence dynamics at each transition frequency across the absorption line, so that variations in the properties of the resonant ions will cause large variations in the shape of the measured excitation spectrum and reveal underlying spectral structure. Specifically, the echo power depends on the material absorption coefficient $\alpha$ and sample length $L$  according to \cite{sun2012a, thiel2012a}
\begin{equation}
I_{echo}\sim [e^{-\alpha L} \textrm{sinh}(\alpha L/2)]^2 e^{\tfrac{-4 t_{12}}{T_2}}, 
\label{echodep1}
\end{equation}
and for the case of small $\alpha L$, as we have for this crystal, Eq. \ref{echodep1} simplifies to
\begin{equation}
I_{echo}\sim [\alpha(\nu)]^2 e^{\tfrac{-4 t_{12}}{T_{2}(\nu)}}. 
\label{echodep2}
\end{equation}
The dependence described by Eq. \eqref{echodep2} provides a powerful tool for interpreting the photon echo excitation spectra. If the inhomogeneous lineshape $\alpha(\nu)$ is already known from direct absorption measurements, this relation allows deviations from the square-root of the observed echo excitation spectrum to be interpreted in terms of variation in optical coherence lifetime $T_2 (\nu)$, identifying subgroups of ions that experience different dynamics. Alternatively, if the coherence lifetime $T_2 (\nu)$ is constant across the inhomogeneous line, i.e. $T_2 (\nu) \equiv T_2$, Eq. \eqref{echodep2} can be used to extract the absorption lineshape from the photon-echo-excitation spectrum.

The measured photon-echo-excitation spectrum is shown in Fig. \ref{fig:lineshape}. We plot the square root of the echo area (i.e. $\sqrt{I_{echo}}$) to provide a direct comparison with the absorption coefficient. The spectrum is smooth and exhibits a Lorentzian lineshape centered at 376.945 THz with a FWHM of 57 GHz, matching the results obtained from spectral hole burning to within experimental accuracy. The decrease in echo excitation signal for the tails of the absorption line is likely a result of the very weak echo signals for frequencies where the absorption coefficient is small but could potentially indicate a slight increase in decoherence for the more strongly perturbed Tm$^{3+}$ ions. Nevertheless, the echo excitation results verify that the decoherence properties are uniform across more than 100 GHz of absorption bandwidth with no “bad” spectral regions due to underlying structure or unresolved defect lines.

\subsection{Magnetic field dependence}

\begin{figure}
\begin{center}
\includegraphics[width=\columnwidth]{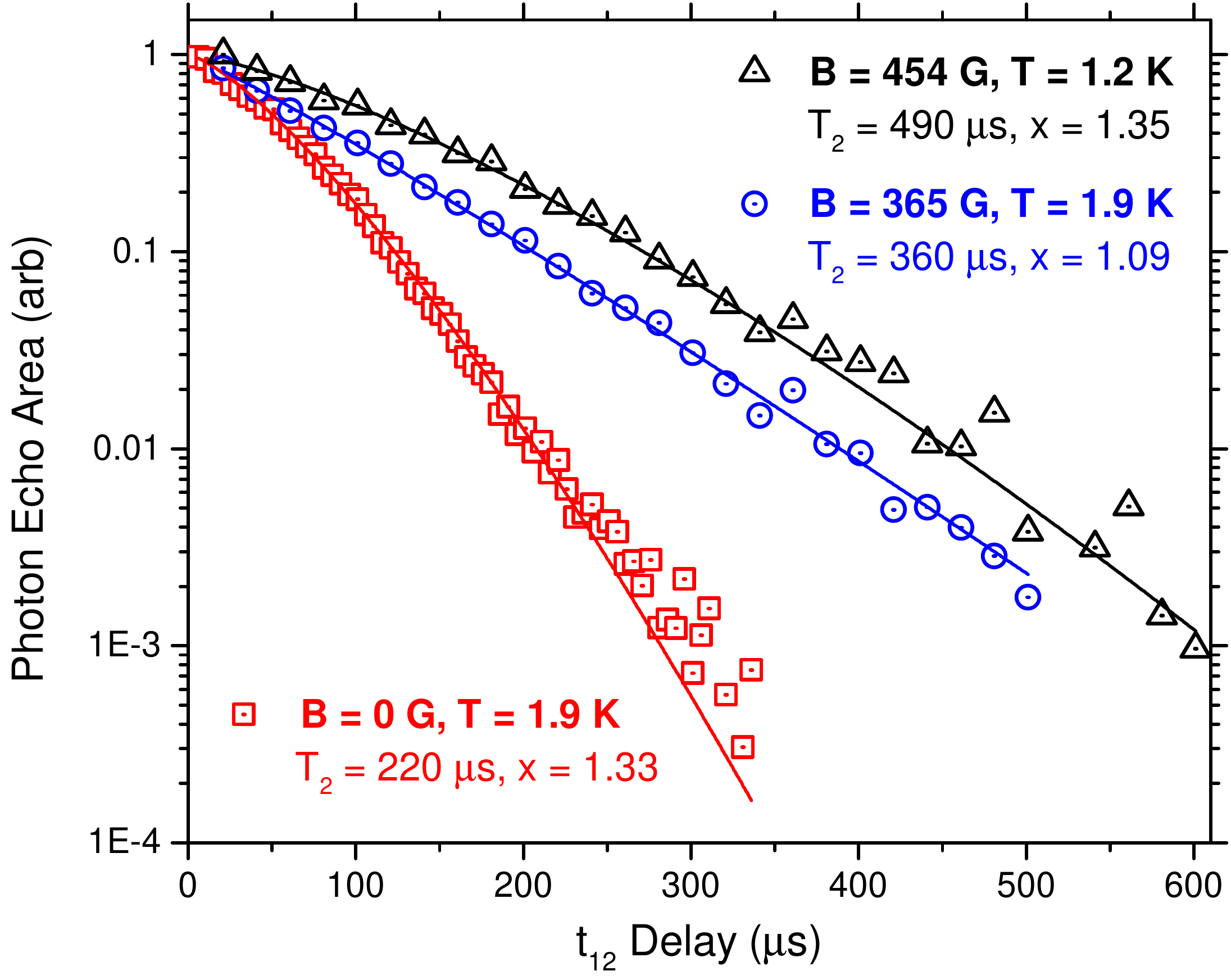}
\caption{Example two-pulse photon echo decays for different temperatures and applied magnetic field strengths. Measurements at 1.9 K in zero field (squares) and with a magnetic field of 365 G (circles) reveal an increase in effective coherence lifetime from 220 $\mu$s to 360 $\mu$s. Further increasing the field to 454 G, reducing the temperature to 1.2 K, and attenuating the laser to reduce ISD-induced broadening yields a coherence lifetime of 490 $\mu$s (triangles). Parameters extracted from each fit (solid lines) are shown.
} 
\label{fig:2pe}
\end{center}
\end{figure}

Next, we measure two pulse photon echo decays at the center of the absorption line (795.325 nm) with and without an applied magnetic field. If spectral diffusion occurs over the timescale of the coherence lifetime due to time-dependent perturbations caused by dynamics in the ions’ environments, the progressive acceleration of decoherence can cause the observed photon echo decay shape to become non-exponential \cite{hu1978a}. A typical photon echo decay at 1.9 K with no applied magnetic field is plotted in Fig. \ref{fig:2pe}, revealing a non-exponential shape that indicates the presence of spectral diffusion in Tm:YGG. The decay is fit using the empirical Mims expression \cite{mims1968a} given by
\begin{equation}
I (t_{12}) = I_0 e^{-2(\tfrac{2 t_{12}}{T_2} )^x},
\label{mims}
\end{equation}
where $t_{12}$ is the delay between the two excitation pulses, $I_0$ is the initial integrated echo intensity at $t_{12}=0$, $T_2$ is the effective coherence lifetime related to the effective homogeneous linewidth $\Gamma_h=1/\pi T_2$ \cite{macfarlane1987a}, and $x$ is an empirical parameter that depends on the magnitude, rate, and nature of the spectral diffusion mechanism. For the zero field data, a $1/e$ coherence lifetime of 220 $\mu$s is determined by fitting Eq. \ref{mims} to the measured decay curve, as shown by the solid red line in Fig. \ref{fig:2pe}. The fit yields a relatively small exponent of $x=1.33$ indicating a modest effect of spectral diffusion on the coherence lifetime under these conditions.

For spectral diffusion arising from fluctuating local magnetic fields caused by nuclear or electron spin flips of host lattice constituents, paramagnetic impurities, or other magnetic defects we expect an applied magnetic field to affect the decoherence dynamics by reducing the magnetic entropy of the spin system. Even at very low strengths, a magnetic field can act to detune different spin-flip transitions and inhibit off-diagonal spin-spin coupling, slowing spectral diffusion from spin flip-flops in the environment and reducing the observed homogeneous linewidth. We find that spectral diffusion in Tm:YGG is significantly reduced by the magnetic field, as shown by the data in Fig. \ref{fig:2pe} for an applied field of 365 G, which results in the coherence lifetime increasing to a value of 360 $\mu$s. Increasing the field to 454 G, reducing the temperature to 1.2 K, and reducing the excitation intensity to minimize broadening due to instantaneous spectral diffusion (ISD) \cite{thiel2014c} results in an even longer coherence lifetime of 490 $\mu$s with $x=1.35$ as shown in Fig. \ref{fig:2pe} \cite{thiel2014a}. The decay exhibits a non-exponential shape, indicating that the coherence lifetime is still limited by spectral diffusion in this low-temperature case. Since spectral diffusion depends on temperature, magnetic field strength and orientation, and perhaps even crystal quality, we expect that optimization of the system could further increase the coherence lifetime.

\subsection{Temperature dependence}

Because different decoherence mechanisms each exhibit a characteristic temperature dependence, studying the variation in homogeneous linewidth with temperature can provide insight into the material dynamics that cause optical decoherence. Quantifying the relationship between coherence lifetime and temperature is important for understanding the physical processes limiting the coherence lifetimes and therefore for predicting potential performance under new conditions. For Tm:YGG, we expect elastic Raman scattering of phonons \cite{mccumber1982a} to be the dominant decoherence mechanism at low temperatures since the large crystal field splittings of 70 cm$^{-1}$ and 26 cm$^{-1}$, in the ground- and excited-state multiplets respectively \cite{sun2005a}, minimize other direct phonon interactions for temperatures below 10 K. In some materials, an additional broadening component can be present due to thermally activated low-energy dynamic structural fluctuations, often described using the theory of two-level systems (TLS) developed for amorphous solids \cite{anderson1972a,phillips1972a}; this results in a quasi-linear term in the temperature dependent linewidth \cite{flinn1994a,macfarlane2004a}. It is known that a low density of TLS may appear even in single crystals if large random lattice strains are present \cite{watson1995a,macfarlane2004a}. Because of the large inhomogeneous broadening in Tm:YGG as well as the observation of significant strain birefringence (see Fig. \ref{fig:birefringence}), we include TLS as a potential decoherence mechanism in our analysis of the observed behavior. As a result, we expect the homogeneous linewidth to be described by
\begin{equation}
\Gamma_h (T) = \Gamma_0 + \alpha_{TLS} T^{\beta} + \alpha_R T^7,
\label{temp}
\end{equation}
where $\Gamma_0$ is the linewidth at zero Kelvin. In the analysis of our measurements, we employ the common approximation of setting the TLS exponent $\beta$ equal to one since the TLS contribution to the linewidth is too weak to unambiguously determine the exponent over the measured temperature range.

\begin{figure}
\begin{center}
\includegraphics[width=\columnwidth]{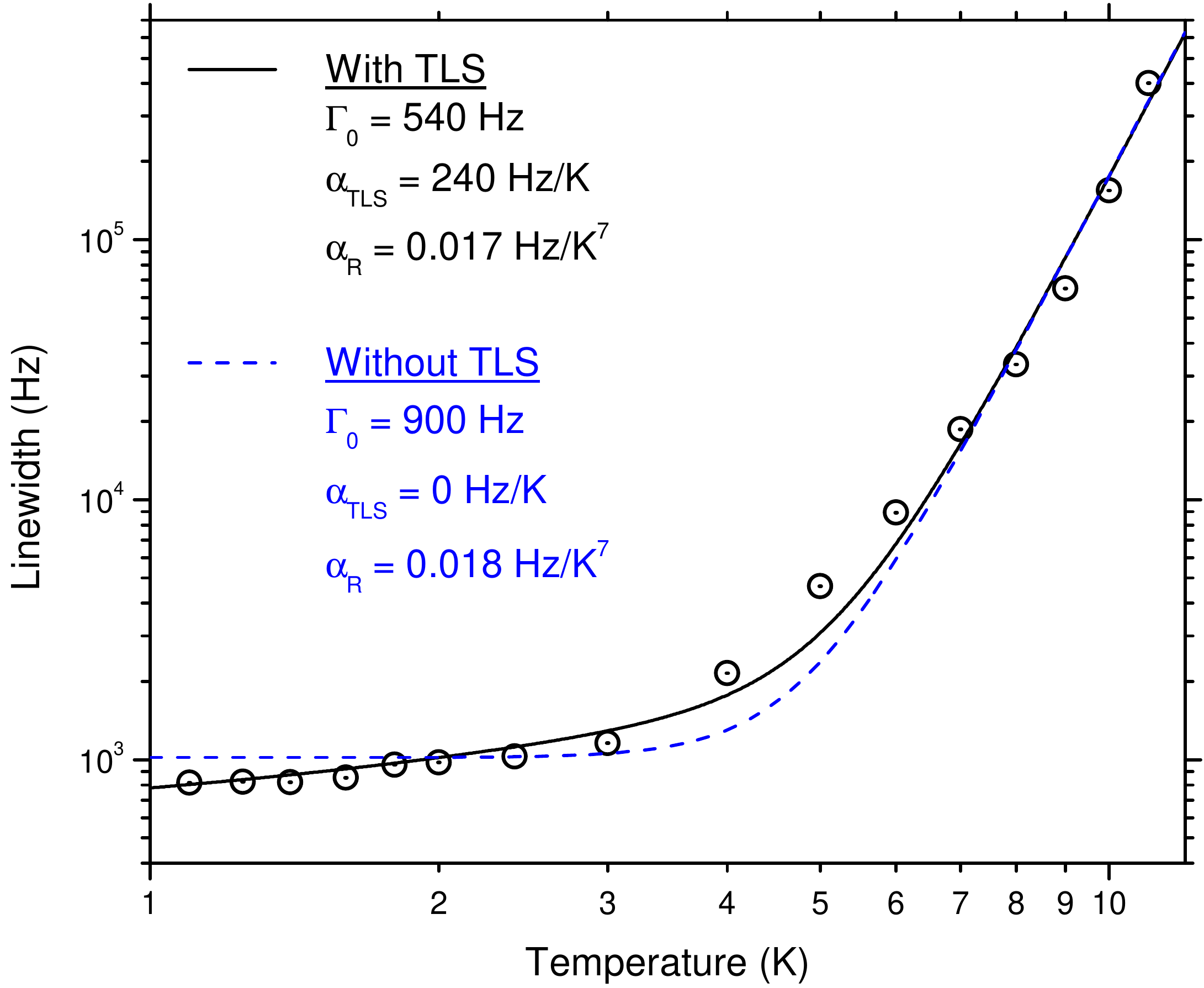}
\caption{Temperature dependence of the homogeneous linewidth in zero magnetic field exhibiting effects of elastic phonon scattering above 4 K and a weak linear broadening below 4 K that may indicate the presence of decoherence due to TLS  in the lattice. Parameters extracted from fits with (solid line) and without (dashed line) TLS included in the model are shown.
} 
\label{fig:temp}
\end{center}
\end{figure}

We measure the variation in coherence lifetime using two-pulse photon echoes and plot the corresponding homogeneous linewidths on a log-log scale in Fig. \ref{fig:temp}. For temperatures above 4 K, a rapid increase in linewidth which is characteristic of the Raman phonon scattering process is observed. At lower temperatures, the dependence cannot be described by either Raman or any other potential direct phonon mechanisms. This much weaker dependence on temperature could indicate the presence of decoherence due to TLS disorder modes in the lattice. The fit of Eq. \eqref{temp} to the data is shown by the solid line, with good agreement over the entire measured temperature range with values of $\Gamma_0 =$ 540 Hz, $\alpha_R =$ 0.017 Hz/K$^7$, and $\alpha_{TLS}$ = 240 Hz/K. For comparison we also plot the best fit of Eq. \eqref{temp} without the TLS component ($\alpha_{TLS} =$ 0). Another mechanism that could potentially exhibit a linear temperature dependence is decoherence due to spin flips of paramagnetic chemical impurities in the crystal \cite{sabinsky1970a}; however, that mechanism would also exhibit a significant magnetic field dependence that we do not observe when comparing decoherence observed at 454 G and 6.4 kG (see Fig. \ref{fig:SDhighB}), suggesting that TLS are the more probable source of decoherence. Since the thermal broadening below 4 K is relatively weak, further measurements are required to definitively verify the presence of broadening due to TLS in this system.

\subsection{Time-dependent spectral diffusion broadening}

Finally, we study the time evolution of spectral diffusion-induced decoherence using stimulated photon echo measurements. Stimulated photon echoes are produced by creating a population grating in the inhomogeneous line using a pair of excitation pulses separated by $t_{12}$, and then scattering a third excitation pulse from the spectral grating after a time delay of $t_{23}$. Just as the temperature dependence provides insights into the material dynamics, the time dependence of spectral diffusion broadening can be used to identify different decoherence mechanisms. Spectral diffusion due to a single class of dilute point perturbers, such as paramagnetic impurities, produces an increase in linewidth with time that approaches a maximum broadening of $\Gamma_{SD}$ at characteristic rate $R_{SD}$, with the parameter values determined by the details of the mechanisms \cite{mims1968a,bottger2006a,klauder1962a,mims1972a}. In contrast, the theoretical models for spectral diffusion due to a distribution of low-energy TLS modes predict a logarithmic increase in linewidth with $t_{23}$ \cite{black1977a,breinl1984a,littau1992a,silbey1996a}. Incorporating both mechanisms into the model for the time evolution of the linewidth, the spectral diffusion broadening can be described by the relation
\begin{equation}
\Gamma_h (t_{23})= \Gamma_0 + \gamma_{TLS} \textrm{ log}(\frac{t_{23}}{t_0}) + \frac{1}{2} \Gamma_{SD} (1-e^{-R_{SD} t_{23}}),
\label{SD}
\end{equation}
where $\Gamma_0$ is the homogeneous linewidth at zero delay, $\gamma_{TLS}$ is the TLS coupling coefficient, and $t_0$ corresponds to the minimum measurement timescale ($t_0 =$ 1 $\mu$s for our experiment).

\begin{figure}
\begin{center}
\includegraphics[width=\columnwidth]{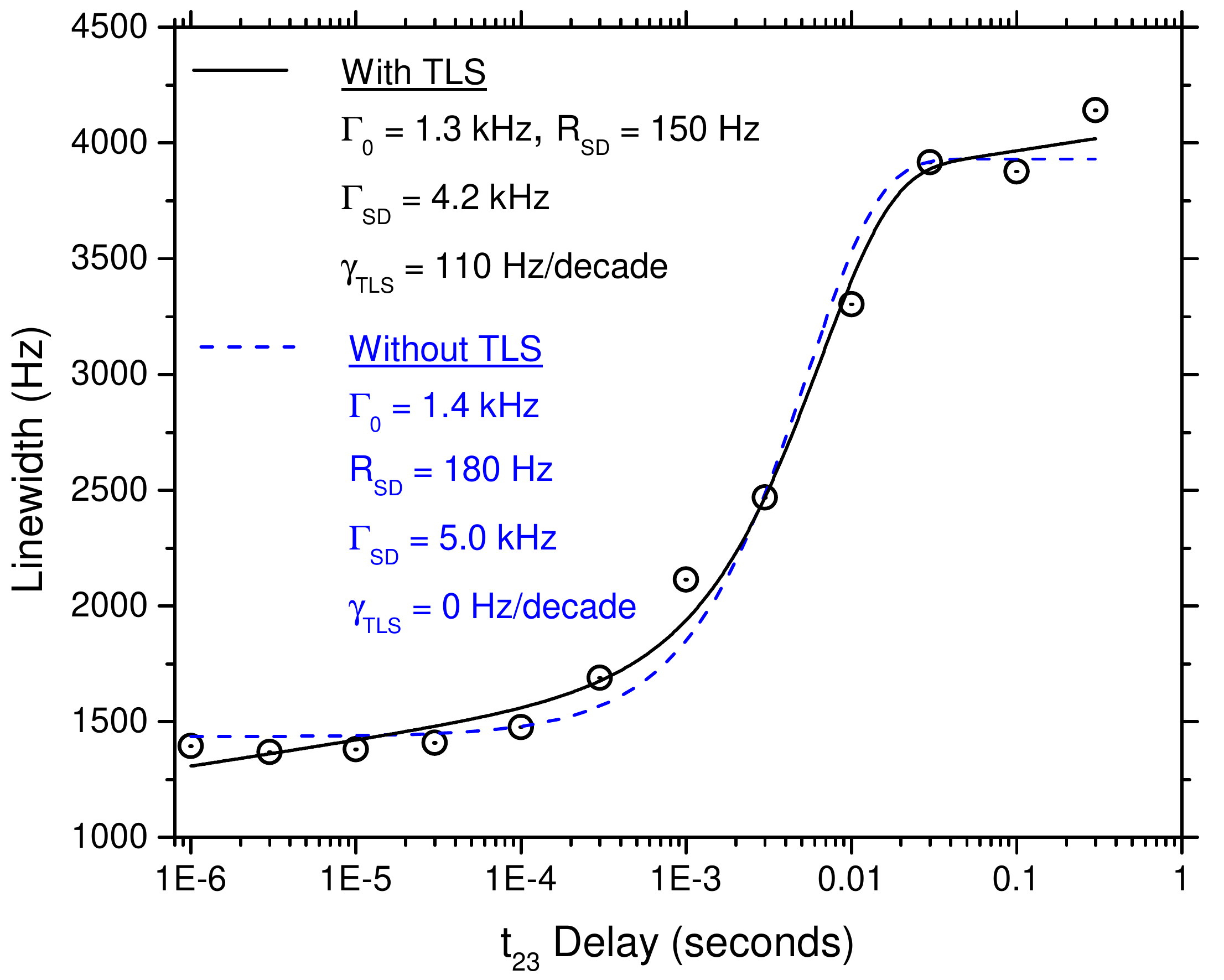}
\caption{Time evolution of the effective homogeneous linewidth at 2.0 K and 358 G revealing the presence of spectral diffusion likely due to fluctuating $^{69}$Ga and $^{71}$Ga nuclear spins. Parameters extracted from fits with (solid) and without (dashed) additional spectral diffusion due to TLS are shown.
} 
\label{fig:SD}
\end{center}
\end{figure}

We measure the variation in effective homogeneous linewidth using stimulated echoes as a function of $t_{23}$ at 2.0 K and 358 G and plot the results in Fig. \ref{fig:SD}. The dependence exhibits the characteristic shape of spectral diffusion broadening due to a single class of perturbers, such as nuclear spin flips in the host lattice \cite{klauder1962a,mims1972a}. The fit of Eq. \ref{SD} without including TLS ($\gamma_{TLS} = 0$) is shown by the dotted line, indicating 5.0 kHz of spectral diffusion acting with a rate of 180 Hz. Based on the shape and magnitude of the spectral diffusion, it seems likely that the dominant source is nuclear spin flips of neighboring gallium in the lattice. To investigate the possibility that TLS contribute to the spectral diffusion, we fit the full expression in Eq. \ref{SD} to the data, as shown by the solid line in Fig. \ref{fig:SD}. Inclusion of the TLS term only results in a slight improvement in the fit of the model to the data; as a result, more detailed studies over a wider range of parameters would be required to unambiguously determine whether TLS contribute to the spectral diffusion time-dependence over these measurement timescales.

\begin{figure}
\begin{center}
\includegraphics[width=\columnwidth]{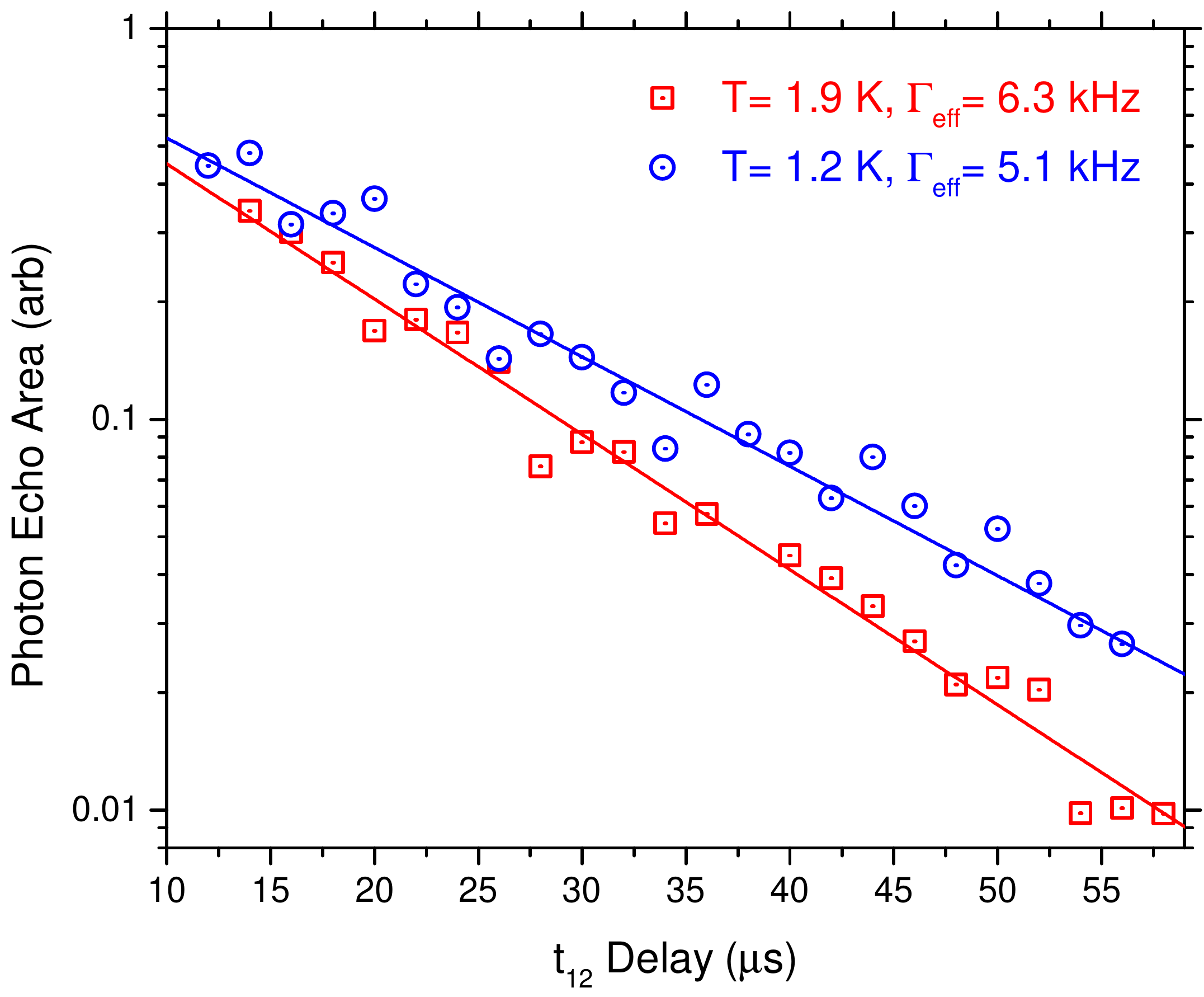}
\caption{Stimulated photon echo decays with a delay of $t_{23}$ = 30 ms and a magnetic field of 6.4 kG revealing a small reduction in effective homogeneous linewidth as the temperature is reduced from 1.9 K (squares) to 1.2 K (circles).
} 
\label{fig:SDhighB}
\end{center}
\end{figure}

To determine whether any further increase in magnetic field strength reduces the impact of spectral diffusion, stimulated photon echo decay measurements are performed under a higher magnetic field strength of 6.4 kG. Photon echo decays are measured with a fixed waiting time of $t_{23} =$ 30 ms at which the spectral diffusion broadening reached the maximum value in the measurements at low fields. Results from measurements at temperatures of 1.9 K and 1.2 K are plotted in Fig. \ref{fig:SDhighB}. Fitting each curve to Eq. \eqref{mims} gives $x=0$ and coherence lifetimes of $\sim$ 60 $\mu$s corresponding to effective homogeneous linewidths of several kHz. These values are comparable to the measurements at lower field, indicating that the additional order-of-magnitude increase in magnetic field strength does not have a significant effect on the spectral diffusion broadening, a result consistent with decoherence due to nuclear spin flips alone or in combination with potential TLS disorder modes in the lattice.

\section{Projection of Additional Improvements}

Using the data presented above, we may make rough projections of improved decoherence properties that may be achievable. At lower temperatures of 0.5 K, we combine the estimated thermal broadening due to TLS of 240 Hz/K from the analysis shown in Fig. \ref{fig:temp} with the zero excitation linewidth of 580 Hz at 1.2 K determined from an ISD analysis on Tm:YGG discussed in Ref. \cite{thiel2014c}. Together, these results predict that the homogeneous linewidth could be as narrow as 410 Hz at 0.5 K for the current crystal, corresponding to a coherence lifetime of 780 $\mu$s. Furthermore, if an improved crystal were to be grown without the TLS that appear to contribute to the linewidth, then the current results suggest that the homogeneous linewidth could be as narrow as 290 Hz, corresponding to a 1.1 ms coherence lifetime. All of these projections, however, have to be considered in view of the nuclear-spin-induced spectral diffusion that is a limiting factor at timescales greater than milliseconds. Therefore, to exploit such long coherence lifetimes under current conditions, applications must operate in less than the spectral diffusion timescales. For example, to increase the storage times of quantum memories based on Tm:YGG, further spectroscopic studies in conjunction with other methods such as coherent population manipulation or optical stimulation methods should be explored.

\section{Conclusion}

Our measurements show that Tm:YGG offers both (i) uniform coherence properties across the broad 56 GHz-wide inhomogeneous line and (ii) coherence lifetimes that are significantly longer than other known Tm-doped materials. Thus Tm:YGG is a superior candidate for broadband and spectrally multiplexed photonic applications.

We measured the magnetic field and temperature dependence of coherence lifetimes with values increasing up to 490 $\mu$s with a magnetic field of 454 Gauss and a temperature of 1.2 K. Although the spectral-diffusion-induced decoherence does not appear to be reduced by further increases in magnetic field strength, employing lower temperature or different crystal orientations could result in further reduction. It is instructive to compare the present results for Tm:YGG to the measured spectral diffusion in 0.1\%Tm:YAG for similar temperatures and magnetic field strengths. In Tm:YAG, the observed spectral diffusion is caused by $^{27}$Al nuclear spin flips is described by values of $R_{SD}$ = 140 Hz and $\Gamma_{SD}$ = 41 kHz \cite{thiel2013a}. For Tm:YGG, we find that the spectral diffusion has a similar rate but with nearly an order of magnitude weaker broadening, a larger difference than would be expected by only considering the relative magnitude of the aluminum and gallium nuclear magnetic moments. Moreover, there are indications that weak dynamic disorder modes in the crystal may be responsible for some of the decoherence observed at the lowest temperatures. A comparison of material properties from different growths or garnet compositions may suggest improved crystal growth or treatment strategies to reduce or eliminate the effects of dynamic disorder modes. 

\section{Acknowledgements}

The authors acknowledge support from Alberta Innovates Technology Futures (AITF), the National Engineering and Research Council of Canada (NSERC), the Defense Advanced Research Projects Agency (DARPA) Quiness program (contract no. W31P4Q-13-l-0004), and the National Science Foundation (NSF) award nos. PHY-1212462 and PHY-1415628.  Any opinions, findings and conclusions or recommendations expressed in this material are those of the authors and do not necessarily reflect the views of DARPA or NSF. W.T. is a senior fellow of the Canadian Institute for Advanced Research (CIFAR).

\end{document}